\batchmode
\documentstyle[12pt]{article}
\begin{document}

\baselineskip=22pt plus 0.2pt minus 0.2pt
\lineskip=22pt plus 0.2pt minus 0.2pt

\vspace*{0.2in}

\begin{center}
{\huge
Aharonov-Bohm problem for spin-$1$\\}

\vspace*{0.25in}
{\Large
 M.\ L.\ Horner and Alfred S.\ Goldhaber\footnote{{\it email}
goldhab@insti.physics.sunysb.edu}}

\vspace*{0.25in}
{\baselineskip=14pt
Department of Applied Mathematics and Theoretical
Physics\\
 University of Cambridge, Silver Street, Cambridge, CB3
9EW, UK}

\vspace*{0.25in}
{\baselineskip=14pt
Institute for Theoretical Physics,
 State University of New York\\
Stony Brook, NY 11794-3840, USA [Permanent Address]}

{\large

\vspace{0.25in}

12 February 1997
\vspace{.25in}

ABSTRACT}

\end{center}

The basic AB {\it problem} is to determine how an unshielded tube of
magnetic flux
$\Phi$ affects arbitrarily long-wavelength charged particles impinging on
it. For spin-$1$ at almost all $\Phi$ the particles do not penetrate the
tube, so the interaction essentially
is periodic in $\Phi$ (AB {\it effect}).
Below-threshold bound
states move freely only along the tube axis, and consequent induced
vacuum currents supplement rather than screen $\Phi$.  For a
pure magnetic interaction the
tube must be broader than the particle Compton wavelength, i.e.,
only the nonrelativistic spin-$1$ AB problem exists.

\newpage
\noindent {\bf I. Introduction}

Aharonov and Bohm, in their first paper on the effect which has come to
bear their name \cite{ab59,ES}, also introduced  a novel problem in
quantum physics.  The AB {\it effect} is the set of measurable phenomena
which occur for charged particles unable to penetrate an endless tube of
magnetic flux -- all observables are periodic in the flux
(with period h/$q$,where $q$ is the charge).  What may properly be called
the AB {\it problem} is determining the influence on charged particles of
an {\underline {unshielded}} flux tube, in the limit where the particle 
de Broglie wavelength goes to infinity.  AB \cite{ab59} observed that in
this limit spin-$0$
particles do not penetrate the tube, so that conditions for the AB
effect are satisfied automaticallly.  Later work showed that there are
no particle states bound inside the tube, and that the
inability to penetrate still holds if the Compton wavelength is
long compared to the de Broglie wavelength, i.e., the particle motion is
relativistic.  Further, there are induced vacuum currents outside the
tube, generating an extra flux which screens the total flux towards the
nearest integer number of flux quanta $N$h$/q$ \cite{serebryanyi85}. 

For the spin-$\frac {1}{2}$ Dirac case there are interesting changes.
Now particles are able to penetrate just enough to be
sensitive to the sign of the flux \cite{hel}.  
This fact is connected with the
existence of threshold bound states for electrons whose magnetic
moment is aligned with the flux:   If there are
$N$ whole quanta of flux, then in the $2+1$ dimensional problem obtained
by factoring out motion in the direction along the tube there are $N$
particle states (with magnetic moment parallel to the flux) confined
inside the tube
\cite{ac79}.  In the full $3+1$ dimensional problem, each such state
corresponds to a distinguishable particle with exactly the free electron
mass, and able to move only along the tube.  If there is an additional
fractional flux there is a `quasi-bound' state, or equivalently
 a
phase shift $\frac{\pi}{2}$ (with respect to the corresponding spin-$0$
case)
at threshold for  exactly one
partial wave.  The perfect `impedance match' between the
infinite-wavelength external wave and the internal state at exactly
threshold energy is what permits this minimal nontrivial coupling between
the flux and the outside particles beyond that implied by the AB effect. 
In the spin-$\frac {1}{2}$ case, induced 
external vacuum currents screen the
magnitude of the flux down towards the nearest smaller integer
\cite{gornicki90}, again
showing dependence on the sign of the flux as well as its fractional
part.

The aim of this work is to determine the corresponding answers to the
AB problem for spin-$1$ Yang-Mills particles.  We find that, except
for a discrete set of flux values, there is no penetration by
threshold-energy particles impinging on the tube.  In this sense the
situation resembles that for spin-$0$, where the AB effect holds
exactly.  However, now there is a set of below-threshold bound states,
somewhat more numerous than the threshold bound states for  spin-$\frac
{1}{2}$.  The most dramatic change is that, to have a pure magnetic
field and no other forces affecting an incident particle, the tube must 
be
broader than a vector boson Compton wavelength, so that there is no
relativistic AB problem for spin-$1$. Finally, spin-$1$ vacuum currents
enhance the given flux, a kind of anti-screening familiar from
discussions of QCD and asymptotic freedom in the domain where only
magnetic fields are considered
\cite{nielsen81}. Many qualitative and even quantitative results for the
AB problem may be illuminated by the study of charged
classical particles interacting with a narrow flux
tube \cite {HG}:  The reason is that the role of h, the quantum of
action, often may be played by another quantity with the same dimensions,
$q\Phi$, the product of particle charge with magnetic flux.  

The rest of the paper is organized as follows.  In Sec. II we
explain why the Yang-Mills equation is the appropriate
analogue for spin-$1$ to
the Klein-Gordon equation for spin-$0$ and the Dirac
equation for spin-$\frac{1}{2}$.  In Sec. III we discuss the scattering
solutions for long wavelength, and find  that
the waves do not penetrate the flux, except for a discrete set of flux
values.  In Sec. IV we study the bound state solutions of the linearized
equations, and find that the number of bound states is somewhat greater
than the number of flux quanta. In Sec. V we analyze the behavior of the
vacuum in the presence of a flux line, taking into account the crucial
contributions to the classical Yang-Mills action quartic in the vector
boson fields. In Sec. VI we find insignificant changes
in the analysis if the magnetic field inside the tube is nonuniform.

\noindent {\bf II. Choice of equation}

  Developments of recent decades make the
linearized Yang-Mills equation
the obvious choice to describe electromagnetic interactions of charged
vector bosons:
By now the
successes of the standard model for electroweak interactions and
quantum chromodynamics for strong interactions show that nonabelian
gauge invariance not only is
attractive esthetically but also is utilized by nature. Here lies
the  difference between our approach to the
spin-$1$  problem and that of
Hagen and Ramaswamy \cite{hagram90},
who adopt the Proca equation, generalized by introducing
minimal electromagnetic coupling into what originally was an equation for
neutral vector bosons. The most general linear spin-$1$ equation
\cite{oow} links the
gyromagnetic ratio
$g$ to an electric quadrupole coupling proportional to $g-2$. 
HR's assumptions give $g=1$ and hence a nonzero
 quadrupole coupling, the origin of
divergent high
energy behavior which would preclude 
perturbative renormalizability \cite{Gl59}.
Like the Dirac equation, the Yang-Mills equation implies
$g=2$, hence an exact lock between precession of
spin and momentum in a uniform, static magnetic field.   
The YM choice manifests a
symmetry of charged-particle motion in pure magnetic fields: In classical
physics, particle trajectories 
depend on momentum, but not energy
(which enters only in determining the speed at which any trajectory is
traversed). 
The symmetry insures consistency between the spatial dependences
of YM wave functions  in
the relativistic regime and in what
HR call the Galilean regime, while for HR electric
quadrupole coupling breaks this connection.  
Further, only for
the YM case does the covariant divergence of the (four-vector) spin wave
function vanish, sustaining the physical interpretation of
the wave function as a purely spatial three-vector in the instantaneous
rest frame of the charged particle.  

HR's quadrupole coupling produces
such pathological behavior in very strong magnetic
fields that they require scattering functions not to
penetrate the flux, making the relativistic AB problem
trivial by fiat.  We on the other hand find that very
strong pure magnetic fields acting on charged vector bosons cannot occur,
so that for physical reasons there is no relativistic AB problem.  In
what they call the Galilean limit, HR neglect the
${\cal O} (\frac {1}{M^2})$ electric quadrupole coupling, obtaining a well
defined problem, but with $g=1$ rather than the preferred value $g=2$, and
scattering resembling that for spin-$\frac{1}{2}$, instead of spin-$0$ as
we find.

\noindent {\bf III. Linearized wave equation and threshold-energy
scattering}

In the Yang-Mills equations, the electromagnetic vector potential
$A_{\mu}$ is identified with the $I_{3}=0$ part of the field, and the
positively and negatively charged fields $P_{\alpha}$ and $N_{\alpha}$
are identified with the $I_{3}=\pm1$ parts, where $I_{3}$ is
the third component of the isospin.  The equations may be
written
\begin{equation} \label{ym}
\left[{\cal D}_{\alpha},\left[{\cal
D}^{\alpha},{\cal D}_{\beta}\right]\right]=0,
\end{equation}
with
\begin{eqnarray}
{\cal D}_{\alpha}=\partial_{\alpha}-iqV_{\alpha},\\ \nonumber
V_{\alpha}=T_{+}P_{\alpha}+T_{-}N_{\alpha}+T_{3}A_{\alpha}\\ \nonumber
\left[T_{i},T_{j}\right]=i\epsilon_{ijk}T_{k},\\ \nonumber
T_{\pm }=T_{1}\pm iT_{2}, \nonumber
\end{eqnarray}
and q the charge of
the particle.  Greek indices run over space and time; Roman over space
only.  From here on except where indicated explicitly, we use units
with $\hbar =c = 1.$
The positive charge projection (all terms with net
unit positive charge)
of (\ref{ym}) contains terms of the form AAP and PNP.  The
latter may be omitted to obtain an equation linear in the charged field.

For perturbative renormalizability the Higgs
mechanism
 is needed to describe masses of vector bosons.
In the linearized wave equation this is functionally equivalent to
 adding a term with a fixed mass $M$, so that the solutions
$P^{\alpha}$ automatically obey the condition
\begin{equation} \label{bg}
D_{\alpha}P^{\alpha}=0,
\end{equation}
with
$D_{\alpha}=\partial_{\alpha}-iqA_{\alpha}$ \cite{oow}.
  There occur in
(\ref{ym})
 two terms of the form $[D_{\alpha},D_{\beta}]$.   Recognizing
$-i\epsilon_{ijk}=S_{k},$
one finds a magnetic moment interaction proportional to $qsB$
(where
$s=\pm 1,0$ is the eigenvalue of ${\bf S \cdot \hat B}$ acting on $P$).
The resulting equation is
\begin{equation}
\label{weq} D_{\beta}D^{\beta}P^{i}+2qsBP^{i}+M^2P^{i}=0.
\end{equation}

To solve (\ref{weq}), we choose the applied magnetic flux in the form
of a uniform cylinder in the z-direction, with radius $R$ taken to
zero at the end of the calculation.
Later we shall come back to the significance and generality of
conclusions associated with assuming uniform field
inside the tube.  Because of the translational and boost symmetries in
the
z-direction we may restrict our analysis to
the two transverse spatial dimensions.

The
(external) kinetic
energy is assumed small in comparison with the magnetic moment
interaction inside the flux tube,  and so
is dropped.  For a state localized well within the flux tube,
the squared wave number $k^2 = E^2 - M^2$ is
given by the following expression, in which the first
term corresponds to the Landau level energy and the second corresponds to
the magnetic moment interaction:
\begin{equation}
k^2 =
\frac{4F}{R^{2}}(n+\frac{1}{2})-\frac{2F}{R^{2}}gs.
\end{equation}
Here the flux $F$ is measured in units of an AB quantum of
the conventional flux $\Phi$, i.e., $F=q\Phi/2\pi$.
For spin-$1$ particles
with $g=2$, this expression can be
negative only for
the lowest Landau level.
Both inside
and outside the flux tube the wave function may be expressed as
\begin{equation} \label{p}
P(r,\phi)=e^{im\phi}f(r),
\end{equation}
where f(r) tacitly depends on the spin projection $s$ and also on the
integer azimuthal angular momentum
$m$, which must be an integer for the wave function to be single-valued. 
Putting (\ref{p}) and the cylindrical forms of the derivatives into
(\ref{weq}) yields
\begin{equation} \label{feqin}
f''+\frac{1}{r}f'-\left[\left(\frac{m}{r}\right)^{2}-qB(m+2s)+
\left(\frac{qBr}{2}\right)^{2}\right]f=0
\end{equation}
inside the flux cylinder, and
\begin{equation} \label{feqout}
f''+\frac{1}{r}f'-
\left[\left(\frac{m}{r}\right)^{2}-qBm\left(\frac{R}{r}\right)^{2}+
\left(\frac{qB}{2}\left(\frac{R^{2}}{r}\right)\right)^{2}-k^2\right]f=0
\end{equation} outside.

The exterior (Bessel) equation is independent of the spin.  Its solution
is:
\begin{equation} \label{ext}
f(r)=cJ_{|m-F|}(kr)+dY_{|m-F|}(kr),
\end{equation}
where J and Y are
respectively the regular and irregular Bessel functions,
and again $F=qBR^{2}/2$
is the number of flux quanta.  The interior solution
may be approximated by a series expansion,
\begin{eqnarray} \label{int}
f(r)=e^{-\frac{Fr^{2}}{2R^{2}}}\left(\frac{r}{R}\right)^{|m|}\times
\ \ \ \ \ \ \ \ \ \ \ \ \ \ \ \ \ \ 
\ \ \ \ \ \ \ \ \ \ \ \ \ \ \ \ \ \ \\
\left  (1+\frac{\Gamma(1+|m|)}{\Gamma(\frac{1}{2}(1-m+|m|-2s))}
\sum^{\infty}_{j=1}\left[(F)^{j}\left(\frac{r}{R}\right)^{2j}
\frac{\Gamma(j+\frac{1}{2}(1-m+|m|-2s))}{\Gamma(j+1)\Gamma(j+
|m|+1)}\right]
\right)\! .
\nonumber \end{eqnarray}
Note that this form is an asymptotic series, since the
radial dependence of the coefficients in the differential equation
precludes analyticity.  Thus care is required in drawing quantitative
conclusions from the use of this approximation, but it should be
good enough for qualitative insight, as it
exhibits the appropriate `antigaussian' asymptotic behavior --
growth at large r given by $e^{+qBr^2/4}$.
In all the following, we shall insure that sufficient
accuracy is available for the purposes at hand.

Now
the inside and outside solutions must be matched at the flux boundary.
The azimuthally dependent factors  and their derivatives match already,
so only radial matching conditions are needed.  We use a two-step
matching that simplifies the bookkeeping.  Near the flux tube and for
small enough values of its radius, the external solution may be written as
\begin{equation} \label{close}
f(r)=a\left(\frac{r}{R}\right)^{|m-F|}+b\left
(\frac{r}{R}\right)^{-|m-F|}.
\end{equation}
The relationship between the coefficients in (\ref{ext})
and those in (\ref{close}) is obtained by expanding (\ref{ext}) 
(using standard asymptotic formulae
for Bessel functions of small argument \cite{gradryzh})
and setting this equal to (\ref{close}).  At the boundary R, the
dimensionless quantities
\begin{equation}
{\bf D} = R\frac{f'}{f}
\end{equation}
 for (\ref{int}), and for (\ref{close}) must match.
This  means $c/d$ must satisfy the equation
\begin{equation} \label{match}
\frac{|m-F|+{\bf D\/}}{|m-F|-{\bf D\/}}=
-\left(\frac{kR}{2}\right)^{2|m-F|}
\frac{\Gamma(1-|m-F|)\left(c+d\frac{\cos(|m-F|\pi)}{\sin(|m-F|\pi)}\right)}
{\Gamma(1+|m-F|)\left(d\frac{1}{\sin(|m-F|\pi)}\right)}.
\end{equation}
The relationship between c and d determines the behavior of
the wave function at large values of the argument (kr).
For $k\geq 0$, the phase shift is defined by
\begin{equation}
tan(\delta)=-d/c.
\end{equation}
Note that we define $\delta$ in such a way that it
would vanish if the charged particle were excluded from the
flux tube.  Of course there is still an AB centrifugal potential,
which means that there is a phase shift from the case of no
flux, but that effect is well-understood; it is the possibility
of deviations from the pure AB case which 
we are trying to address here.  
The behavior of $\delta$ as a
function of F is given by (\ref{match}).  For less than critical values of
the flux, $\delta$ is small and positive.  At the critical value of F,
$\delta$
rises sharply through $\pi /2$ to just below $\pi$, where it remains for
larger than critical flux.  The size of $kR$ determines how sharp the
transition is.  For $kR=0$, $\delta(F)$ is a step function. The
transition  from
a free to a bound state occurs when (\ref{match}) can be satisfied for
$k=0$.  At precisely this value of $F$, a quasi-bound state exists, 
i.e., $d/c$ diverges as $k$
approaches zero.  Such a wave function
 has an infinitely long tail, so that it is
not square integrable, but for infinitesimally larger $F$
it would be a true bound state.  According to (\ref{match}) the
quasi-bound state occurs for $F$ such that
\begin{equation} \label{cond}
|m-F|+{\it\bf D\/}=0,
\end{equation}
of course
only possible when the magnetic moment
 and the flux are parallel.  The quasi-bound state with the smallest flux
occurs for
$m=1$ at $F=0.74$.  Note that the seeming
solution of (\ref{cond}) $F=0$ is
spurious. Instead, for
$m=0$ and any $F\ne 0$ there is a true bound state, the
more deeply bound the bigger $|F|$ is.

Just
as in the case of spin-$\frac {1}{2}$
 (and in HR's Galilean limit of the Proca scheme for spin-$1$), where
quasi-bound states exist for all noninteger $F$, the
existence of such a state implies penetration of the flux tube by the
particle, sufficient to produce sensitivity to the sign of the flux.  The
difference for spin-$1$ Yang-Mills particles
is that the quasi-bound states exist only for
discrete values of the flux, so that penetration occurs only for flux
values in a set of measure zero.

\noindent {\bf IV. Counting bound states}

At energies less than the mass, i.e., $k^2<0$, the matching conditions
yield (\ref{match}) with $k$ replaced by $i\kappa$.
Since these are bound states
and not just quasi-bound, the large $r$ behavior must be a decaying
exponential, which means $c/d=1/i$.  Then $F$ must satisfy
\begin{equation} \label{boundcond}
\frac{|m-F|+{\bf D\/}}{|m-F|-{\bf D\/}}=-(\frac{\kappa
R}{2})^{2|m-F|}\frac{\Gamma(1-|m-F|)}{\Gamma(1+|m-F|)}.
\end{equation}
For $\kappa=0$, this also dictates that $F$ satisfy (\ref{cond}):
One has approached the quasi-bound-state limit from the bound-state
rather than the scattering side, but the limiting behavior is the same.
Therefore a value of $F$ greater
than a critical value by even the smallest amount implies the existence of
a true bound state
in the corresponding partial wave.
As mentioned earlier, despite the fact that for $m=0$ there is never a
quasi-bound state, a true bound state does exist for any
non-zero value of $F$.  To count the
total number of bound states we need
to find that value of $F$ for which a quasi-bound state
appears at a given $m$; any $F$ slightly greater than this yields exactly
$m+1$ bound states.

The dependence of the total number of bound states on $F$ can be
inferred at least roughly
from the approximate solutions of (\ref{cond}).  For
each increase of $m$ by one, the number of
possible bound states increases by one.
Therefore, the change in $F$ per added bound state at some value of $F$
can
be found by solving (\ref{cond}) for pairs of adjacent values of $m$.  We
fit a curve to points obtained this way, using the approximation
(\ref{int}), and sought to obtain an asymptotic form for $\frac{dm}{dF}$.
Integrating the resulting expression gave
an estimate for the number of bound states $\nu$
as a function of the amount of flux,
\begin{equation} \label{numbound}
\nu \approx F+0.3\sqrt{F}.
\end{equation}

Because we know that the series method is not quantitatively reliable,
this result needs further examination.  First, it is worth noting that
the qualitative character of (\ref{numbound}) is quite reasonable.
Since states in the lowest Landau level all are bound, there should be
at least $[F + 1]$ of them.  Near the edge of the tube, there should be
extra room for some less strongly bound configurations, and the number
of these should be proportional to the circumference of the tube
$2\pi R$,
measured on the scale of the magnetic length, $R\sqrt{\pi/F}$.  Thus
simple geometry underlies this extra contribution to the number of
bound states.

In the large $F$ limit an asymptotic form for $m_{max}$ as a function
of
$F$
(where $m_{max}$ is the maximum azimuthal quantum number corresponding
to a bound state)
can be found by writing (\ref{weq}) in
the form
\begin{equation} \label{approx}
\frac{d^{2}f(r)}{dr^2}+\frac{1}{r}\frac{df(r)}{dr}+k^{2}(r)f(r)=0,
\end{equation}
where
\begin{equation}
k^{2}(r)=\frac{4F}{R^2}-\left(\frac{m-F\left(\frac{r}{R}\right)^2}
{r}\right)^2.
\end{equation}
We expand $k(r\equiv \bar R-\delta)$
through
second order about its minimum
at $r=\bar R = R\sqrt{m/F}$ (even though we know
$\bar{R}>R,$ the outer radius of the tube), and make the
substitution
\begin{equation}
x=\sqrt{F}\delta/R.
\end{equation}
If we assume that $m$ can be written as
\begin{equation} \label{scale}
m=F+\alpha \sqrt{F},
\end{equation}
then, recalling the assumption that
F is very large, (\ref{approx}) becomes
\begin{equation} \label{scaled}
\frac{d^{2}\tilde{f}(x)}{dx^2}+4(1-x^2)\tilde{f}(x)=0.
\end{equation}
Matching logarithmic derivatives across the flux
tube boundary results in an equation for $\alpha:$
\begin{equation}
\frac{\tilde{f'}\left(\frac{\alpha}{2}\right)}{\tilde{f}\left(
\frac{\alpha}{2}\right)}=\alpha,
\end{equation}
where the radius $R$ of the flux tube
corresponds to $x=\alpha/2$.  A
direct numerical solution of (\ref{scaled}) converged well and gave
\begin{equation} \label{coef}
\alpha=0.55.
\end{equation}
To this same two-place accuracy, the JWKB approximation
carried consistently through second order in $x$ gives the same result,
which is rather impressive, as the inside-outside
matching condition is imposed not
far from the classical turning point where the approximation has a
spurious square root divergence.

\noindent {\bf V. Vacuum polarization effects}

Having counted the bound state solutions to the linear wave equation,
we need to analyze their effect on the vacuum structure.  If the flux
is spread out on a scale large compared to the boson Compton
wavelength, then the bound states have positive energy smaller than the
rest mass.  Clearly this lowers the vacuum energy compared to that in
the absence of the flux, and therefore produces a paramagnetic effect
enhancing that flux, a clear example of antiscreening.  The
antiscreening may compete with, but should dominate,
 the effect of threshold scattering states,
which as for spin-$0$ tend to bring the flux to the nearest quantum value,
whether larger or smaller in magnitude.

A quite different situation arises if the flux is assumed to be
concentrated so that the magnetic length is less than the Compton
wavelength.  In this case, the bound states have $\omega^2 = -\kappa^2
+m^2 < 0$, so that the frequency is imaginary, and the bound state
amplitudes grow exponentially with time.  This is not vacuum polarization,
but rather instability of what one would naively identify as the vacuum.
The first thing one can say is that this instability
must be halted by
the terms in the energy quartic in the charged boson field, which 
act as an effective mass proportional to the field amplitude,
and eventually must counterbalance the negative quadratic 
terms responsible for
the instability.  It is an interesting question worth further study
whether the configuration obtained by optimizing the
coefficients of the unstable modes of the linearized equation is itself
stable, or whether additional instabilities bring about the complete
extinction of the entire Yang-Mills field strength inside a very narrow
tube.

There are
several reasons to believe that this might be the case.  First, on
distance scales small compared to the boson Compton wavelength the full
nonabelian gauge invariance is manifest, and the flux, which is a gauge
covariant rather than invariant quantity, should not be a physical
observable with a definite nonzero expectation value.  Secondly, if we
try to imagine how this flux could be created, it would require a
cylindrical sheet of intense current.  The gauge interaction of the
particles producing this current would generate huge quantum fluctuations
in the isospin orientation of each particle, so that its charge would
average to zero, as would the corresponding current.  Hence there would
be no steady source for the flux, and so no flux.  Finally, Nielsen and
Olesen
\cite{NO79}
observed that a vacuum instability in QCD which favors
formation of a uniform
nonzero magnetic field does not by itself end in a
stable configuration.   There
is a further instability to formation of what they call flux
spaghetti, i.e., a complicated pattern of tubes of flux rapidly
varying in space and time.  This suggests that a single isolated flux tube
of very small radius cannot occur.  Either there are none, or there
are many tending to cancel each other.  Here we find the most dramatic
change from the situations for lower spin.  There is nothing inconsistent
about a flux line influencing spin-$0$ or spin-$\frac{1}{2}$.  However,
for spin-$1$,  the unadorned flux-line concept only makes sense if the
particles are
nonrelativistic, so that
the flux tube size can be bigger than the boson
Compton wavelength, yet still smaller
than the de Broglie wavelength.

The above conclusion is consistent with known models for flux tubes
in relativistic field theories.  In these models, the tubes are
examples of cosmic strings, with finite energy per unit length.  The
radius of such a string is determined by a force balance which
automatically precludes magnetic lengths smaller than the vector boson
Compton wavelength \cite {VS}.

\noindent {\bf VI. Nonuniform field distributions}

	We promised to consider cases where the magnetic field is not
uniform inside the tube.  A nonuniformity involving magnetic
length scales smaller than the Compton wavelength appears unphysical,
for the reasons just discussed. Otherwise, the conclusion for the
uniform-field case should continue to hold, that except for
flux configurations
in a set of measure zero where quasi-bound
states occur,
the scattering solutions at large de Broglie wavelength
do not
penetrate the flux.  For bound states, the situation could be more
complicated.  For example, suppose that there were many `islands' of flux,
each carrying a positive flux $F_i < 0.74$.  Provided there were sufficient
spacing between islands compared to the radius of any one, each island
would have one bound state, and the total number of bound states for
large total $F$ would be proportional to $F$ but with a proportionality
constant $1/F_i > 1$.  If field of both signs is allowed, then the number
of bound states $N$ could exceed the net flux $F$ by an arbitrarily large
factor, but the difference between the numbers of spin-up and of
spin-down bound states would be more closely linked to $F$.  This
statement actually applies also to the spin-$\frac {1}{2}$ case, where
there is an exact index theorem, $N_{up}-N_{down}=[F]$
\cite{ac79}.  There the
exterior behavior, i.e., the finite-energy scattering, depends
only on the sign of the total flux and on its fractional part.

For all three spins the low-energy scattering on a flux tube is
determined by the fractional part of the flux, and for spin-$\frac {1}{2}$
also the sign.  For spin-$0$ there are no bound states regardless of the
distribution of the flux, but for the higher spins the number of bound
states is sensitive to the distribution.  Thus one finds the unsurprising
conclusion that behavior inside the flux tube depends on the
distribution, but behavior outside is completely unaffected, except
for configurations in a set of measure zero in the case of
spin-$1$.  In other words, exterior sensitivity to the flux distribution,
as opposed to the total flux,
shows little or no change with spin, precisely because there is
little or no penetration of the flux.

\noindent {\bf VII. Conclusions -- Spin metamorphoses of the
Aharonov-Bohm problem}

The problem of a charged particle in the presence of a flux line
originated with the paper of Aharonov and Bohm \cite{ab59}, where they
observed that in the absence of spin the
particle automatically is excluded from the flux.  Thus all phenomena
must be periodic in the flux, with a period of one AB flux quantum.  For
spin-$\frac{1}{2}$ there are exactly
$[F]$ normalizable zero-energy states bound inside the flux,
and also one quasi-bound state,
as long as $F$ exceeds its integer part
$[F]$ by
any nonzero amount \cite{ac79}.  It is this feature which allows
the wave function to penetrate the flux just enough to be sensitive to
its sign, thus slightly spoiling the perfect AB periodicity of the
spinless case, and violating usual expectations for decoupling between
phenomena at very different scales.
Nevertheless, the problem of spin-$\frac{1}{2}$ particles interacting
with an arbitrarily thin flux tube remains
well-defined.  For spin-$1$, the linearized Yang-Mills equation has
clear solutions in the limit of zero tube radius, but the bound state
solutions growing with time (which appear if the limit is taken on the
scale of the Compton wavelength of the vector boson) are
physically unacceptable.  Fortunately, the nonlinearities
in the Yang-Mills system conspire to make this limit unachievable.  On
length scales large compared to the Compton wavelength the limit does make
sense, and the description of scattering and the counting of bound states
go through exactly as described in the body of the paper.  Such a
`fat' flux line would polarize the vacuum so as to enhance the
applied flux.  As the tube radius cannot be made small compared to
the Compton wavelength,
the spin-$1$ case appears to be the end of
the  road for the relativistic Aharonov-Bohm problem, though the more
complex problem which replaces it deserves further study.

\vspace*{0.25in}

\begin{center}
\large

ACKNOWLEDGMENTS

\end{center}

This work was supported by the Patricia Harris Fellowship, the National
Science Foundation, and the Particle Physics and Astronomy Research
Council.  Martin Bucher independently performed the numerical
integration leading to the result (\ref{coef}) for the number of
bound states. Martin Ro\v cek made instructive comments about
nonabelian gauge effects.

\noindent 

\end{document}